\documentclass[aps,12pt,superscriptaddress,preprint,floatfix]{revtex4-1}
\usepackage{amssymb,graphics,graphicx}
\usepackage{amsmath,amssymb,amsthm,bm,multirow,longtable}
\renewcommand{\epsilon}{\varepsilon}
\renewcommand{\phi}{\varphi}

\theoremstyle{plain}

\theoremstyle{definition}
\allowdisplaybreaks

\begin{document}
\title{Sum Rule Constraint on Models Beyond the Standard Model}
\author{Paul H. Frampton}
\email{paul.h.frampton@gmail.com}
\affiliation{
Courtyard Hotel, Whalley Avenue, New Haven, CT 06511, USA}
\author{Thomas W. Kephart}
\email{tom.kephart@gmail.com}
\affiliation{Department of Physics and Astronomy, Vanderbilt University, Nashville, TN 37235}
\date{\today}

\begin{abstract}
In most versions of beyond the standard model (BSM) physics, the Yukawa couplings of the quarks and
charged leptons are not all to the same complex scalar doublet but to 
different ones. Comparison to the standard model (SM) with only one scalar doublet,
using the known mass of the W boson, provides a sum rule
constraint on the Yukawa couplings $Y_i,~~ i=t, b, \tau,.....$ of the form $\Sigma_i  r_i^2 = 1$
where $r_i = Y_i^{(SM)}/Y_i^{(BSM)}$ and the sum is over 
distinct scalar doublets.
The LHC data on the branching ratios $H \rightarrow \gamma\gamma,
~\bar{b}b,  ~\tau^+\tau^-, ~etc.,$ allows detailed comparison to this sum rule constraint
and, as accuracy improves, will constrain or exclude many BSM theories.
\end{abstract}
\maketitle

Starting with the discovery, in 2012,
of the scalar boson H with mass $M_H \simeq 126$ GeV \cite{CMSandATLAS} and appropriate CP and spin properties \cite{CP,spin}
underlying the Brout-Englert-Higgs mechanism \cite{Higgs:1964ia,Englert:1964et} for spontaneously breaking \cite{Guralnik:1964eu} the electroweak $SU(2) \times U(1)$ gauge symmetry, we are now entering
a golden age of particle phenomenology, a field which had previously 
been data-starved
for a very long time. In particular, the detailed examination of the properties of H \cite{PDG,Aad:2013wqa,Chatrchyan:2013mxa,couplings}
can drastically whittle down viable possibilities for constructing theories which go
beyond the standard model.

\bigskip

A general characteristic of most models beyond the standard model (BSM) which
distinguishes them from the standard model (SM) is that they contain more than one
complex scalar doublet. The different flavors of quarks and leptons couple
generically not all to the same scalar doublet but to different ones. The detailed
pattern of these couplings varies from model to model but we shall take a general
approach which includes all possibilities. Namely, we shall first assume that each
flavor couples to a different doublet, and then special cases will be degenerate
examples of this general case. In the SM, all flavors couple to the same
scalar doublet.

\bigskip

The Large Hadron Collider (LHC) has not only made the dramatic discovery of the H boson and finally
nailed down its previously-unknown mass but equally importantly opens up
the experimental measurement of the detailed couplings of H through its
production cross section and especially through its decay modes and 
partial decay widths. Of special interest here are the couplings of H to
fermions. We recall the scandal of the fermion masses that none of
the twelve quark and lepton masses have a satisfactory theoretical
understanding. These masses are simply parametrized in the SM
by Yukawa couplings $Y_i$ where $i = t, b, \tau, ...$.

\bigskip

Let us begin by reviewing the situation in the SM. We shall focus
on the third generation fermions $t, b$, and $\tau$ but the generalization
to the lighter fermions will be straightforward. The third generation is the most
relevant to the LHC experiments.

\bigskip

The corresponding Yukawa couplings of the SM are written
\begin{equation}
{\cal L}_Y^{(SM)} =  \left[ Y_t ^{(SM)} \bar{t}t  + Y_b^{(SM)}  \bar{b}b + Y_{\tau}^{(SM)} \bar{\tau}\tau \right] H + c.c.
\label{SMYukawa}
\end{equation}
in terms of the mass eigenstates.
The spontaneous breaking occurs through the BEH mechanism where H develops
a vacuum expectation value $<H>$ uniformly throughout the universe and given by
\begin{equation}
<H> = V = (\sqrt{2} G_F)^{-1/2} \simeq 246 ~ {\rm GeV}.  
\end{equation}

\bigskip

From Eq.(\ref{SMYukawa}), the SM Yukawa couplings 
\begin{equation}
Y_i^{(SM)} = \left( \frac{M_i}{V} \right)
\label{SMYuk}
\end{equation}
for $i=t, b, \tau$.
 have the values
$Y_t^{(SM)}   \simeq 0.704$, 
$Y_b^{(SM)}  \simeq 0.0170$,
 and
$Y_{\tau}^{(SM)}  \simeq 0.00722$,
 where we have used $M_t=173.07$ GeV, $M_b = 4.18$ GeV, and $M_{\tau} =1.77682$ GeV.

\bigskip

Note that the W mass $M_W$ is given by
\begin{equation}
M_W^2 = \left( \frac{g_2^2 V^2}{4} \right) = (80.385 {\rm GeV})^2
\label{Wmass}
\end{equation}
where $g_2$ is the gauge coupling for the $SU(2)$ factor of the electroweak gauge group.

\bigskip
\bigskip
\bigskip

In a BSM model, the generalization of Eq.(\ref{SMYukawa}) involves different H doublet
scalar fields and can be written
\begin{equation}
{\cal L}_Y^{(BSM)} =  Y_t ^{(BSM)} \bar{t}t H_t + Y_b^{(BSM)}  \bar{b}b H_b  + Y_{\tau}^{(BSM)} \bar{\tau}\tau H_{\tau}  + c.c.
\label{BSMYukawa}
\end{equation}
and, writing the VEVs as $<H_i> = V_i$,  the generalization of Eq.(\ref{SMYuk}) are now written in the form
\begin{equation}
Y_i^{(BSM)} = \left( \frac{M_i}{V_i} \right)
\label{BSMYuk}
\end{equation}
for $i=t, b, \tau$.

\bigskip

In such a theory, the W mass is given by a generalization of Eq.(\ref{Wmass}) to

\begin{equation}
M_W^2 = \left( \frac{g_2^2}{4} \right) \Sigma_i V_i^2 = (80.385 {\rm GeV})^2
\label{BSMWmass}
\end{equation}
where the sum is over the distinct scalar doublets, {\it i.e.},   any of the $H_i$ fields in
Eq. (\ref{BSMYukawa}) that are identified separately, are included in the sum only once.

\bigskip

Defining 
\begin{equation}
r_i = \left( \frac{Y_i^{(SM)}}{Y_i^{(BSM)}} \right)
\label{rdefinition}
\end{equation}
then using Eqs.(\ref{SMYuk},\ref{Wmass},\ref{BSMYuk},\ref{BSMWmass})
one finds the useful sum rule
\begin{equation}
\Sigma_i r_i^2 = 1
\label{sumrule}
\end{equation}
where, in any given BSM model, the summation is restricted as discussed following Eq. (\ref{BSMWmass}). Note that there could, in principle, be further scalar doublets
$K_i$ with $<K_i> =k_i \neq 0$ coupling normally to $W$ but not at all to $t, b,$ and $\tau$ whereupon
Eq.(\ref{sumrule}) is $\Sigma_i r_i^2 \leq 1$. However, because the unequality is not experimentally motivated and, in any case, serves only to strengthen all the constraints discussed, we shall 
focus on an equality sign in Eq.(\ref{sumrule}).
                                                    
\bigskip

One interesting consequence of the sum rule, Eq. (\ref{sumrule}), is that consistency with experiment
requires that
\begin{equation}
|Y_i^{(BSM)}| \geq |Y_i^{(SM)}|
\label{lowerbound}
\end{equation}
for all $i = t, b, \tau, ...$

\bigskip

There exist a large number of BSM theories in the literature and a majority
of the popular ones fall into one of two classes, (I) and (II), as follows 
proceeding in a direction away from the standard model:

\bigskip

{\bf Class I:}
\noindent{\it
In Eq.(\ref{BSMYukawa}), the $b$ and $\tau$ scalar doublet are identified,
$H_b=H_{\tau}$.}

\bigskip

\noindent
In this class, the sum rule simplifies to
\begin{equation}
r_t^2 + r_b^2 = r_t^2 + r_{\tau}^2 = 1 ~~~ r_b = r_{\tau}
\label{classI}
\end{equation}
and it is conventional to parametrize
\begin{equation}
V_t = V {\rm sin} \beta ~~~ V_b = V_{\tau} = V {\rm cos} \beta
\end{equation}

\bigskip

\noindent
Examples of Class I are the minimal supersymmetric standard model (MSSM),
the most usual type of two Higgs double model (2HDM), and the Peccei-Quinn
model (PQ).

\bigskip

{\bf Class II:}
\noindent
{\it In Eq.(\ref{BSMYukawa}), the scalar doublets $H_t, H_b,, H_{\tau}$ are
all distinct.}

\bigskip

\noindent
In this case, the sum rule is
\begin{equation}
r_t^2 + r_b^2 + r_{\tau}^2 = 1
\label{classII}
\end{equation}
and it is conveneient to parametrize the VEVs as
\begin{equation}
V_t = V {\rm sin} \beta ~~~ V_b = V {\rm cos} \beta ~{\rm sin} \alpha 
~~~ V_{\tau} = V {\rm sin} \beta ~{\rm cos} \alpha
\end{equation}

\bigskip 

\noindent
Most renormalizable flavor models using as symmetry $SU(3) \times SU(2) \times U(1) \times G_F$
where $G_F$ is a global flavor symmetry are of this class.  Many models of this type have appeared in the literature \cite{Altarelli:2010gt,Ishimori:2010au},  including in our own work \cite{Frampton:1994rk}.

There are some BSMs that are not constrained by the sum rule Eq. (\ref{sumrule}). These have extra Higgs doublets, but they do not get VEVs.
For example, inert Higgs models \cite{InertHiggs} can be of this type. See \cite{Arhrib:2014pva} for a recent discussion.

\bigskip
\bigskip

Our purpose here is mainly to present the sum rule constraint Eq. (\ref{sumrule}) on building BSMs, but we now indicate
how one can confront the already existing LHC data with this constraint. Here we give just a few examples and use
the present LHC experimental results to demonstrate the procedure. A more complete analysis will be presented elsewhere.

To  lowest order the amplitude for $H \rightarrow \tau \bar{\tau}$ is
proportional to $Y_{\tau}$ and the dominant production is by gluon fusion via a top loop
so the cross section goes like $(Y_tY_{\tau})^2$.
Likewise to lowest order   the cross section $H \rightarrow b \bar{b}$ goes like $(Y_tY_b)^2$.
The CMS and ATLAS experiments quote values for the cross section and
compares it with that predicted by the SM. (See \cite{Chatrchyan:2014nva,ATLAS-CONF-2013-108} for $H \rightarrow \tau \bar{\tau}$ decays and
\cite{Grippo:2014zea,ATL-PHYS-PUB-2014-011} for $H \rightarrow b \bar{b}$.)

\begin{table}[htdp]
\caption{$r_t^2r_{\tau}^2$ lower limits}
\begin{center}
\begin{tabular}{|c|c|c|c|}
\hline
\hline
LHC Collaboration    &   $1\sigma$  &        $2\sigma$    &       $3\sigma$\\
\hline
\hline
CMS            &                0.952          &    0.719         &      0.581\\
\hline
ATLAS         &                0.562      &       0.400        &       - - - \\
\hline
\hline
\end{tabular}
\end{center}
\label{rtauTable}
\end{table}%

As an example we now use the $1\sigma$ CMS results to extract values for $r_{\tau}$ and $r_b$.
For $m_H = 125$ GeV, the CMS best fit of the observed $H \rightarrow \tau \tau$ signal strength is $(0.78 \pm 0.27)$, which is the ratio of cross section times branching
fraction for BSM to the SM  $Y_{\tau}^{SM} = m_{\tau}/V$ and we use
 $m_{\tau} = (1776.82 \pm 0.16)$ MeV.
{\bf If we ignore  alternative sub-dominant production processes then
we have}
\begin{equation}
\sigma(gg \rightarrow H \rightarrow \tau \tau)_{expt} = (0.78 \pm 0.27) \sigma (gg \rightarrow H \rightarrow \tau\tau)_{SM}   ~~~~~~~~~~~~~~1\sigma~ ({\rm CMS})
\label{sigmatau}
\end{equation}
where all errors quoted are one standard deviation.
We also assume all the difference from the SM is in the Yukawas.
{\bf The total width of the scalar can be altered by such changes but this
is a measurement which can be made independently to confirm or
refute deviations from the SM.} Hence we conclude
$(Y_t^2Y_\tau^2)_{expt} = (0.78 \pm 0.27)~(Y_t^2Y_{\tau}^2)_{SM}$
where we identify $(Y_i)_{expt}$ with the BSM Yukawa $(Y_i)_{BSM}$.
We can also write Eq.(\ref{sigmatau}) as
\begin{equation}
r_t^{-2}r_{ \tau}^{-2} = 0.78 \pm 0.27 ~~~~~~~~~~~~~~~~~~~~~~~~1\sigma~ ({\rm CMS})
\label{rtau}
\end{equation}
and requiring that the $r_i$ are consistent with our sum rule Eq.(\ref{sumrule}) which dictates
that $r_i^2 \leq 1$ or $r_i^{-2} \geq 1$ gives
\begin{equation}
r_{ \tau}^2 \ge 0.95, ~~~~~~~~~~~~~~~~~~~~~~~~1\sigma~ ({\rm CMS})
\label{rtauOneSigma}
\end{equation}
thus $r_{ \tau} \ge 0.976$ and   $\tan \beta \le 0.22$.

For $H \rightarrow b \bar{b}$
the CMS signal cross section times branching
fraction for $m_H = 125$ GeV is $(1.0 \pm 0.5)$ times the standard
model expectation, hence (Using $Y_b^{SM} = m_b/V$ where
$m_b = (4.18 \pm .03)$ GeV in the $\overline{MS}$ scheme.)
$(Y_t^2Y_b^2)_{expt} = {(1.0 \pm 0.5)}(Y_t^2Y_b^2)_{SM}$
where as above we identify $(Y_\tau)_{expt}$ with the BSM Yukawa $(Y_\tau)_{BSM}$.
{\bf Again we assume the only unknown in the cross section are the Yukawa couplings,
by ignoring other production processes and effects of the total width, so that we can also write}  
\begin{equation}
r_t^{-2}r_{b}^{-2} = (1.0 \pm 0.5) ~~~~~~~~~~~~~~~~~~~~~~~~1\sigma~ ({\rm CMS})
\label{rb}
\end{equation}
and requiring that $r_i$ remain consistent with Eq.(\ref{sumrule}) gives
\begin{equation}
r_{b}^2 \ge 0.66 ~~~~~~~~~~~~~~~~~~~~~~~~~ 1\sigma ({\rm CMS})
\label{rbOneSigma}
\end{equation}
thus $r_{b} \ge 0.819$, which corresponds to $\tan \beta \le 0.70$. Hence the bound on $\beta$ from $H \rightarrow b\bar{b}$ is somewhat weaker than from $H \rightarrow \tau \tau$, but they both will impact Class I BSMs including a variety of 2HDMs   \cite{Branco:2011iw,Ferreira:2012nv} and various SUSY models \cite{Belanger:2013xza,Dumont:2013npa} including MSSM. We stress that we have made approximations that can and will be improved, but it is clear that the sum rule constraint will have teeth.
As in the above examples, the $1\sigma$, $2\sigma$ and $3\sigma$ lower limit values of $r_{ \tau}^2$ and $r_{b}^2$ are calculated from CMS and ATLAS data and are quoted in tables \ref{rtauTable} and \ref{rbTable} respectively.

Regarding Class II BSMs, even without including $r^2_t$, it is clear that the combined $3\sigma$ CMS result  ($r^2_{\tau}+r^2_b$) will begin to be able to constrain models of this Class.
 Note that CMS results are more restrictive
than those of ATLAS in this case, as for most of the discussion in this paper.

\begin{table}[htdp]
\caption{$r_t^2r_b^2$ lower limits}
\begin{center}
\begin{tabular}{|c|c|c|c|}
\hline
\hline
LHC Collaboration    &   $1\sigma$  &        $2\sigma$    &       $3\sigma$\\
\hline
\hline
CMS            &                0.658            &      0.476           &         0.286\\
\hline
ATLAS         &                0.909     &        - - -       &       - - - \\
\hline
\hline
\end{tabular}
\end{center}
\label{rbTable}
\end{table}%

 \bigskip
 \bigskip
 \bigskip

Now we proceed to the top Yukawa coupling and $H \rightarrow \gamma \gamma$. 
{\bf For this decay the partial width can be extracted directly at LHC by comparison to
other decays: the production mechanism of $H$ is thus factored out. The effect
of the $\gamma\gamma$ decay on total width is very small because of the
tiny branching ratio.} Since the top is heaver than $M_H/2$ we estimate $Y_t$ using the decay mode $H \rightarrow \gamma \gamma$. There are two one-loop
contributions to $H \rightarrow \gamma \gamma$, a top loop and a $W$ loop. We assume the $W$ loop is known and as in the SM, and the deviation from SM results of the decay width of the Higgs in this channel is all in the top Yukawa $Y_t$. We need to compare the data with the SM calculation   \cite{Ellis:1975ap,Ioffe:1976sd,Shifman:1979eb} which has been recently summarized in \cite{Marciano:2011gm}.

This calculation has a venerable history and was first presented
in 1976 \cite{Ellis:1975ap} in a certain limit, then more generally in 1979
\cite{Shifman:1979eb}. These early results were confirmed much more recently
in 2011 \cite{Marciano:2011gm,Shifman:2011ri} in response to a false criticism
by  \cite{Gastmans:2011ks,Gastmans:2011wh}. We therefore use the following established
formulas from \cite {Marciano:2011gm}, generalized for BSM, where the rate is given by
\begin{equation}
\Gamma(H \rightarrow \gamma \gamma)=|F|^2\left(\frac{\alpha}{4\pi}\right)^2\frac{G_F m_H^3}{8\sqrt{2}\pi}
\label{topwidth}
\end{equation}
 where the function $F$ for BSMs is given by
\begin{equation}
F=F(\beta_W)+\Sigma_f  N_c Q_f^2 r_f^{-1}F(\beta_t)
\label{FMarciano}
\end{equation}
with $\beta_W=\frac{4m_W}{m_H^2}$, $\beta_t=\frac{4m_t}{m_H^2}$.
 The standard model form of $F$ is recovered by setting  $r_f=1, \forall ~{}_f$. 
If we include only the top quark in the sum, with
color factor $N_c=3$ and $Q_t=2/3$, then
\begin{equation}
F=F(\beta_W)+\frac{4}{3} r_t^{-1}F(\beta_t)
\label{FBSM}
\end{equation}
where
\begin{equation}
F_W=2+3\beta+3\beta(2-\beta)f(\beta),
\label{FW}
\end{equation}
\begin{equation}
F_t=-2\beta[1+(1-\beta)f(\beta)]
\label{Ft}
\end{equation}
and
\begin{equation}
f(\beta)=\left[\arcsin(\frac{1}{\sqrt{\beta}})\right]^2.
\label{f}
\end{equation}
This last expression is valid for $\beta > 1$ as is true for both $\beta_W$
and $\beta_t$.

Eq.(\ref{topwidth}) with $r_t=1$, as in the SM, is consistent with the observed $H\rightarrow \gamma\gamma$ rate. In a Class I model, such as the MSSM,
on the other hand, the sums rule constraint together with the
$\bar{\tau}\tau$ final state constrain $r_t \geq 0.111$

\begin{figure}
\includegraphics[width=18.cm]{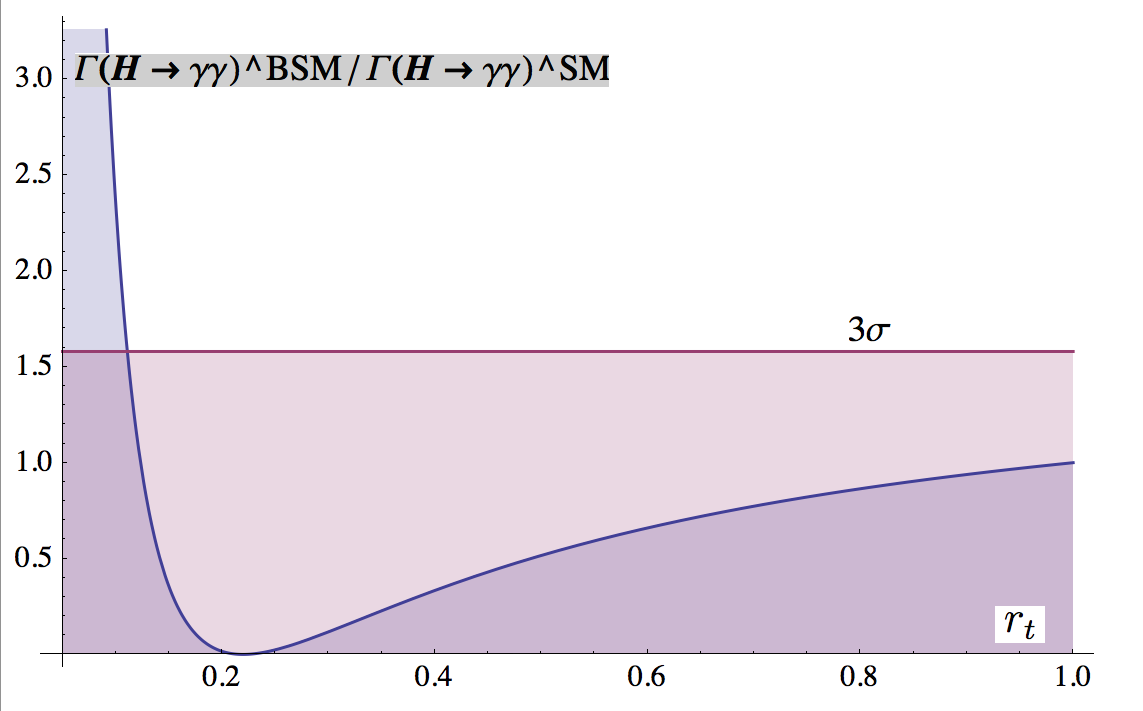}
 \vspace{-1 cm}
\caption{\label{figure1}$\Gamma(H\rightarrow \gamma \gamma)$ for BSM/SM} The ratio of $\Gamma(H\rightarrow \gamma \gamma)$ for the BSM vs the SM as a function of $r_t$. The SM is on the curve at the point (1,1).
\end{figure}

Substituting the observed masses for $W$, $t$ and $H$ we find that,
\begin{equation}
\frac{F_{BSM}}{F_{SM}}= \frac{8.354 - 1.836r_t^{-1} }{6.519}
 \label{FBSM/FSM}
\end{equation}
and thence
\begin{equation}
\frac{\Gamma_{BSM}(H \rightarrow \gamma \gamma)}{\Gamma_{SM}(H \rightarrow \gamma \gamma)}\leq 1.58  ~~~~~~~~~~~~~~~~3\sigma ({\rm CMS})
 \label{GammaBSM/GammaSM}
\end{equation}
which is displayed in Fig. \ref{figure1}.
\begin{table}[htdp]
\caption{Upper limits of measured $H\rightarrow \gamma \gamma$ rate divided by SM rate.}
\begin{center}
\begin{tabular}{|c|c|c|c|}
\hline
 \hline
LHC Collaboration    &   $1\sigma$  &        $2\sigma$    &       $3\sigma$\\
\hline
 \hline
CMS            &                1.04           &      1.31          &         1.58\\
\hline
ATLAS         &                1.88    &        2.21       &       2.54\\
\hline
 \hline
\end{tabular}
\end{center}
\label{gammagammaTable}
\end{table}%
\bigskip
\bigskip

\newpage

The combination of the above results suggests that some MSSM, PQ and
Class II models are disfavored.

The next to leading order (NLO) percentage corrections to the decay width $\Gamma(H\rightarrow \gamma \gamma)$ has been calculated \cite{Passarino:2007fp} where it is found that the electroweak and QCD correction are both about $2\%$ but of opposite signs, so they nearly cancel leading to a  total correction of less than one percent compared with the leading order calculation.

{\bf Additional particles can alter the decay widths, as can the variations of the Yukawa couplings
from their SM values on which we have focused. If the BSM Yukawas $Y_i^{BSM}$ do deviate
from the SM values $Y_i^{SM}$, one may suspect additional states although the range
of possibilities is too wide-ranging to analyze succinctly here. Even if we have focused on $H\rightarrow \bar{b}b, \bar{\tau}\tau, \gamma\gamma$ other
decays such as $ H \rightarrow VV$ where $V$ are vector gauge bosons
can also be useful to probe departure from the SM.} 

Although the preliminary LHC data on H decay is presently of
 limited accuracy, it is nevertheless exciting that it is already
enough to dispose of some examples of BSM models.

With the upcoming second run of the LHC, anticipated to begin
in 2015 at higher energy and luminosity, one can confidently
expect a great improvement in the accuracy of the
measurements for the H partial decay modes and hence a
better and more detailed check of the constraint sum rule.
This heralds a new chapter of particle phenomenology. Constructing
viable theories beyond the standard model will become very
tightly constrained which is obviously a good thing.
There are models with extra Higgs doublets that do not acquire VEVs, like inert Higgs models, that can avoid the sum rule constraint.

To conclude, we have found a sum rule that applies to BSMs that have more than one Higgs doublet with VEVs and  Yukawa coupling to light fermions. The sum rule  constrains all models of this type including but not limited to a large class of flavor symmetry models, 2HDMs, SUSY models including MSSM. \\

\noindent
Acknowledgment: The work of TWK was supported by DoE grant\# DE-SC0011981.

\end{document}